\documentclass[twocolumn,amsmath,amssymb]{revtex4-2}
\usepackage[T1]{fontenc}
\usepackage{amsmath,graphicx,amssymb,epsfig,babel,dsfont,amscd,ulem,tikz}
\usepackage{mathrsfs}
\usepackage{color}
\usepackage{placeins}
\usepackage{hyperref}
\usepackage{physics}
\usepackage{float}
\begin{document}

\title{Hysteresis-Driven Radiative Mpemba Effect in Phase-Change Nanostructures}

\author{F. Herz}
\affiliation{Institut f\"{u}r Physik, Carl von Ossietzky Universit\"{a}t, 26111, Oldenburg, Germany}
\email{florian.herz@uol.de}

\date{\today}

\begin{abstract}
The Mpemba effect states that initially hotter systems cool faster than colder ones. While known in convective, conductive, and quantum systems, its radiative analogue is unexplored. Here, this anomaly is realized via phase-change hysteresis of a VO$_2$ nanoparticle near a SiC substrate. After analytically deriving an onset condition, the phase space is mapped. Crucially, latent heat acts as a thermal buffer enabling both ordinary and inverse effects. Near-field coupling governs the relaxation time and enables a passive effect where memory is stored externally via substrate reflection.
\end{abstract}

\maketitle

%%%%%%%%%%%%%%%%%%%%%%%%%%%%%%%%%%%%%%%%%%%%%%%%%%%%%%%%%%%%%%%%%%%%%%%%%%%%%%%%
%
% Introduction
%
%%%%%%%%%%%%%%%%%%%%%%%%%%%%%%%%%%%%%%%%%%%%%%%%%%%%%%%%%%%%%%%%%%%%%%%%%%%%%%%%

\section{Introduction} \label{Ch:Intro}

The Mpemba effect describes the counterintuitive phenomenon where an initially hotter system cools down faster than an initially colder one, first documented for water \cite{Mpemba1969}. In the last two decades, this anomalous relaxation has been translated to a variety of other systems, including Brownian particles trapped in optical tweezers \cite{Kumar2020}. This platform also enabled the observation of the inverse effect, where an initially colder system heats up faster \cite{Kumar2022}, theoretically explained by the non-monotonic temperature dependence of the maximum extractable work from a local-equilibrium state \cite{Chetrite2021}. In granular fluids, numerical simulations revealed that non-equilibrium statistical velocity distributions can induce the effect \cite{Lasanta2017}. More generally, in Markovian systems without a phase transition, the ruling factor is the second largest eigenvalue of the probability distribution trajectories, explaining why the Mpemba effect occurs in three-state models or antiferromagnetic Ising chains \cite{Lu2017, Ben-Abdallah2026}. 

Very recently, the Mpemba effect was generalized to quantum systems. For instance, a qubit driven by a thermal photon source can undergo the inverse Mpemba effect due to interference \cite{AharonyShapira2024}. In many-body quantum systems, systems whose symmetry is initially more strongly broken can restore it faster \cite{Ares2023}, which was experimentally verified in a trapped-ion quantum simulator \cite{Joshi2024}. Additionally, in a single trapped-ion system prepared in a pure state, this anomalous relaxation was linked to Liouvillian exceptional points \cite{Zhang2025}. For both the classical and quantum Mpemba effect, recent reviews provide comprehensive overviews \cite{Bechhoefer2021, Teza2026}. 

Inspired by these findings, this work translates the Mpemba effect to thermal radiative transport, moving beyond the traditional realms of conduction, convection, or symmetry restoration. While previous work has demonstrated the Mpemba effect in radiative heat transfer for many-body systems via Markovian dynamics~\cite{Ben-Abdallah2026}, the realization presented here exploits non-Markovian memory embedded in phase change hysteresis, enabling anomalous relaxation in a minimal two body geometry. In non-Markovian quantum systems, the Mpemba effect occurs \cite{Strachan2025} due to memory effects, which are also a characteristic of macroscopic complex systems like spin glasses \cite{Baity-Jesi2019}. In thermotronics, which design thermal analogues to electric components, materials undergoing a metal-insulator transition at a critical temperature $T_c$, such as VO$_2$, are widely used for thermal diodes \cite{Yang2013, Ben-Abdallah2013}, transistors \cite{Ben-Abdallah2014, Joulain2015}, and logic gates \cite{Biehs2016, Kathmann2020}. In the transition regime, these phase-change materials exhibit a hysteresis separating the cooling and heating branches. This introduces an intrinsic memory effect and radiative bistability \cite{Kubytskyi2014, Dyakov2015}, which can drive thermal self-oscillations \cite{Dyakov2015}. VO$_2$ is especially promising in the near field because the switch from the metallic ($T>T_c$) to the dielectric phase ($T<T_c$) triggers the excitation of strongly coupling Surface Phonon Polaritons (SPhPs) in the dielectric phase, drastically enhancing the radiative heat transfer compared to the metallic phase \cite{vanZwol2011}. 

The paper is organized as follows. In Sec.~\ref{Ch:Theo}, I introduce the theoretical framework coupling near-field radiative heat transfer with thermal hysteresis via the T-FORC model, and derive the general analytical onset criterion governing the radiative Mpemba effect. In Sec.~\ref{Ch:Res}, I analyze the dynamic interplay between the caloric buffer of latent heat and near-field photon exchange for a phase-change nanoparticle above a metallic substrate. I then demonstrate the ``passive'' Mpemba effect, showing that anomalous relaxation can be driven purely by memory stored externally in the substrate reflection, which is a mechanism unique to the near-field regime. Finally, Sec.~\ref{Ch:Conc} summarizes the key findings.

\section{Theory} \label{Ch:Theo}

\begin{figure}
	\centering
	\includegraphics[width=0.48\textwidth]{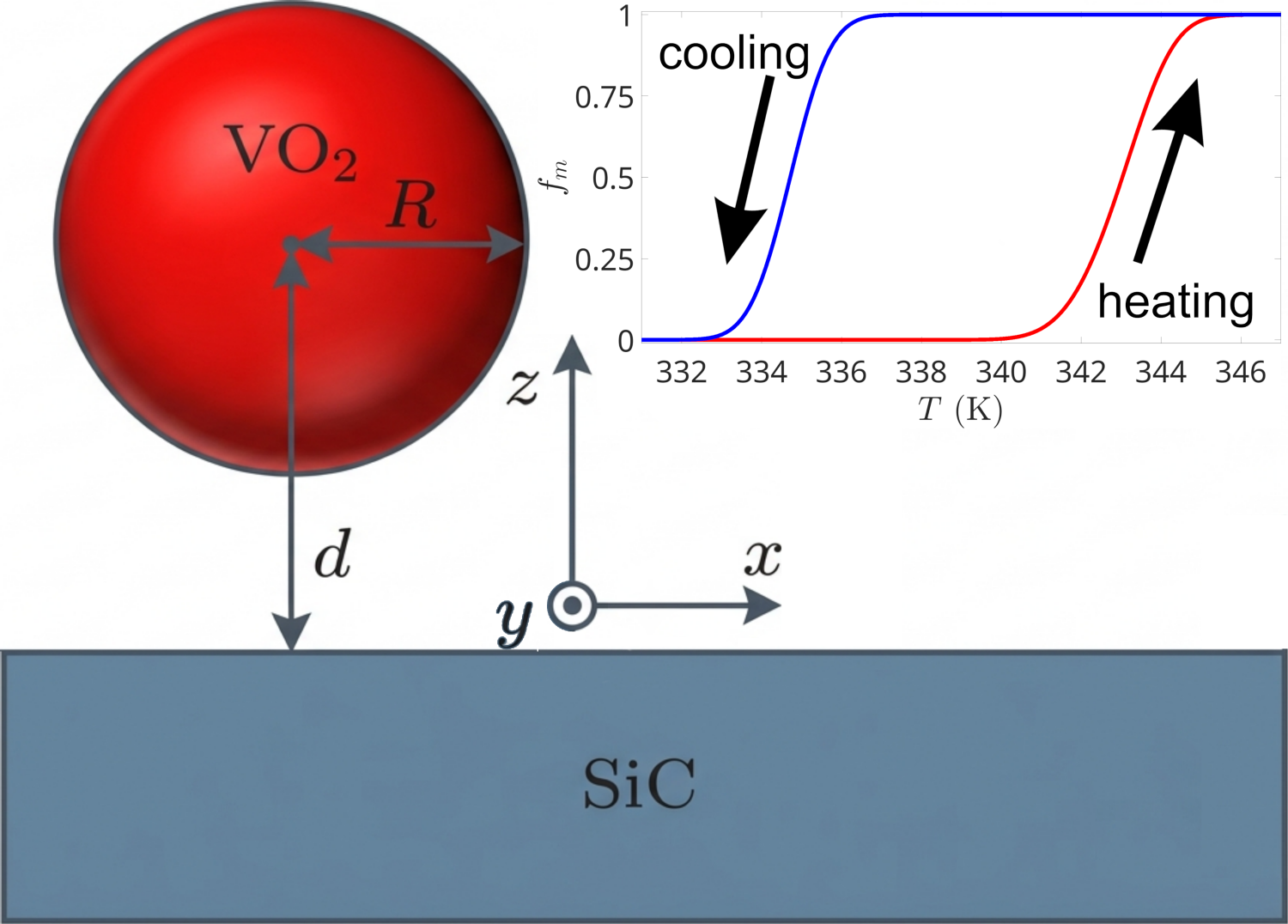}
	\caption{Sketch of the considered system. A VO$_2$ nanoparticle of radius $R$ is placed at center-tosurface distance $d$ above a SiC substrate. The inset shows the hysteresis of the metallic volume fraction $f_m$ in the transition regime 332 K $< T < $ 346 K.}
	\label{Fig:Sketch}
\end{figure}

To demonstrate the emergence of this radiative Mpemba effect, this work analyzes a VO$_2$ nanoparticle of radius $R$ at a center-to-surface distance $d$ above a SiC substrate, which also supports SPhPs, thereby enabling a detailed discussion of near-field radiative effects. This schematic setup is shown in Fig.~\ref{Fig:Sketch}. The substrate's temperature $T_{\rm sub}$ is held constant, while the temperature of the nanoparticle $T_\text{dip}$ evolves.

If the temperature homogenization throughout the body is much faster than the thermal relaxation time, the time evolution of the temperature follows \cite{Incropera_etAl_2007}
\begin{equation}
\frac{\text{d} T_{\rm dip}}{\text{d} t} = - \frac{\mathcal{P}_{{\rm sub} \leftrightarrow {\rm dip}}}{\rho C_{\rm p} V} ,
\label{Eq:Temp_evol}
\end{equation}
where $\rho$ is the mass density and $V$ is the volume. The specific heat capacity $C_{p}$ and the exchanged power $\mathcal{P}_{{\rm sub} \leftrightarrow {\rm dip}}$ are detailed in Sec.~S1 of the Supplemental Material. The calculations of the radiative heat transfer $\mathcal{P}_{{\rm sub} \leftrightarrow {\rm dip}}$ are based on the framework of Fluctuating Electrodynamics \cite{Rytov_Kravtsov_Tatarskii_1989}. To have a Mpemba effect, the time evolution of two distinct trajectories have to cross at a certain time $\tau$ such that $T_1(\tau) = T_2(\tau)$ with $\frac{\text{d} T_1}{\text{d} t} |_{t=\tau} < \frac{\text{d} T_2}{\text{d} t} |_{t=\tau}$ assuming an initially hotter temperature compared to an initially colder one, i.e. $T_1(0) > T_2(0)$. Defining $T_1 = T + \Delta T$ and $T_2 = T$ with $\Delta T(\tau) = 0$, the following relation holds at $t = \tau$
\begin{align}
\frac{\text{d} \Delta T}{\text{d} t} & = \zeta (T + \Delta T, f + \Delta f, \gamma + \Delta \gamma) - \zeta(T, f, \gamma)
\end{align}
with the definition 
\begin{align}
\zeta(T, f, \gamma) & = -\frac{\mathcal{P}_{{\rm sub} \leftrightarrow {\rm dip}} (T, f)}{\rho C_{\rm p} (T, \gamma) V} .
\label{Eq:Temp_evol_1}
\end{align}
Here, $f$ denotes the volume fraction, e.g. the metallic one, responsible for the hysteresis as shown in Fig.~\ref{Fig:Sketch} and $\gamma = \partial f / \partial T$ denotes its time derivative. It stores the memory of whether the system was previously heated or cooled. Therefore, even though $\Delta T = 0$ at $t = \tau$ holds, a structural memory mismatch $\Delta f \neq 0$ and $\Delta \gamma \neq 0$ persists. For small deviations $\Delta T \ll T$, $\Delta f \ll f$, and $\Delta \gamma \ll \gamma$, $\zeta$ can be expanded around the state of the colder trajectory yielding
\begin{align}
\zeta(T + \Delta T, f + \Delta f, \gamma + \Delta \gamma) & \approx \zeta(T, f, \gamma) + \frac{\partial \zeta}{\partial T} \Delta T \notag \\
& \quad + \frac{\partial \zeta}{\partial f} \Delta f + \frac{\partial \zeta}{\partial \gamma} \Delta \gamma.
\label{Eq:zeta_taylor}
\end{align}
Re-inserting this expansion into Eq.~\eqref{Eq:Temp_evol_1} yields
\begin{align}
\frac{\text{d} \Delta T}{\text{d} t} |_{t = \tau} & = \frac{\partial \zeta}{\partial f} \Delta f + \frac{\partial \zeta}{\partial \gamma} \Delta \gamma,
\label{Eq:Temp_evol_2}
\end{align}
confirming that $\Delta f \neq 0$ or $\Delta \gamma \neq 0$ are strictly necessary to observe a Mpemba effect. From Eq.~\eqref{Eq:Temp_evol_2} and the derivatives of $\zeta$ defined in Eq.~\eqref{Eq:Temp_evol_1}, the final condition is obtained as
\begin{align}
& 0 < \frac{\partial \mathcal{P}_{{\rm sub} \leftrightarrow {\rm dip}}}{\partial f} \Delta f - \frac{\mathcal{P}_{{\rm sub} \leftrightarrow {\rm dip}} L}{C_{\rm p}} \Delta \gamma.
\label{Eq:Mpemba_cond}
\end{align}
Note that the derivative $\partial C_{\rm p} / \partial \gamma = L$ is taken from Eq.~(S17) in the Supplemental Material. Hence, the derivatives of the exchanged power and the specific heat capacity define whether an anomalous relaxation occurs. Note that this mathematical derivation holds for any internal variable $f$ whose thermodynamic response introduces a rate-dependent or memory effect in the heat capacity, and is not restricted to hysteresis in phase-change materials. For example, it applies to materials whose optical and caloric properties depend on the heating or cooling rate itself via the fictive temperature ($f = T_f$, $\gamma = \partial T_f / \partial T$) due to structural aging \cite{Narayanaswamy1971, Song2026} or under strain ($f = \sigma$, $\gamma = \partial \sigma / \partial T$) \cite{Zhang2024}, provided that the parameter $L$ is interpreted as the corresponding effective thermodynamic coefficient. An interesting limiting case is the exchange of the materials, placing a SiC particle above a VO$_2$ substrate. In this scenario, $C_{\rm p}$ only depends on $T$ because the memory effect is stored externally in $\mathcal{P}_{{\rm sub} \leftrightarrow {\rm dip}}$ via the substrate's reflection. Then, Eq.~\eqref{Eq:Mpemba_cond} simplifies to 
\begin{align}
& 0 < \frac{\partial \mathcal{P}_{{\rm sub} \leftrightarrow {\rm dip}}}{\partial f} \Delta f.
\label{Eq:Mpemba_cond2}
\end{align}
Because the relaxing particle lacks an intrinsic memory, this scenario is referred to as the ``passive'' Mpemba effect and underscores a key theoretical insight of this work. The emergence of anomalous relaxation does not strictly require an internal state variable within the relaxing body itself. In this passive regime, the body acts as a memoryless probe, while the non-Markovian history is stored externally in the structural state of the substrate. As will be shown, this passive mechanism relies critically on evanescent wave coupling and vanishes in the far-field limit, proving that it is a unique feature of near-field radiative transport.

\section{Results and discussion} \label{Ch:Res}

\begin{figure}
	\centering
	\includegraphics[width=0.48\textwidth]{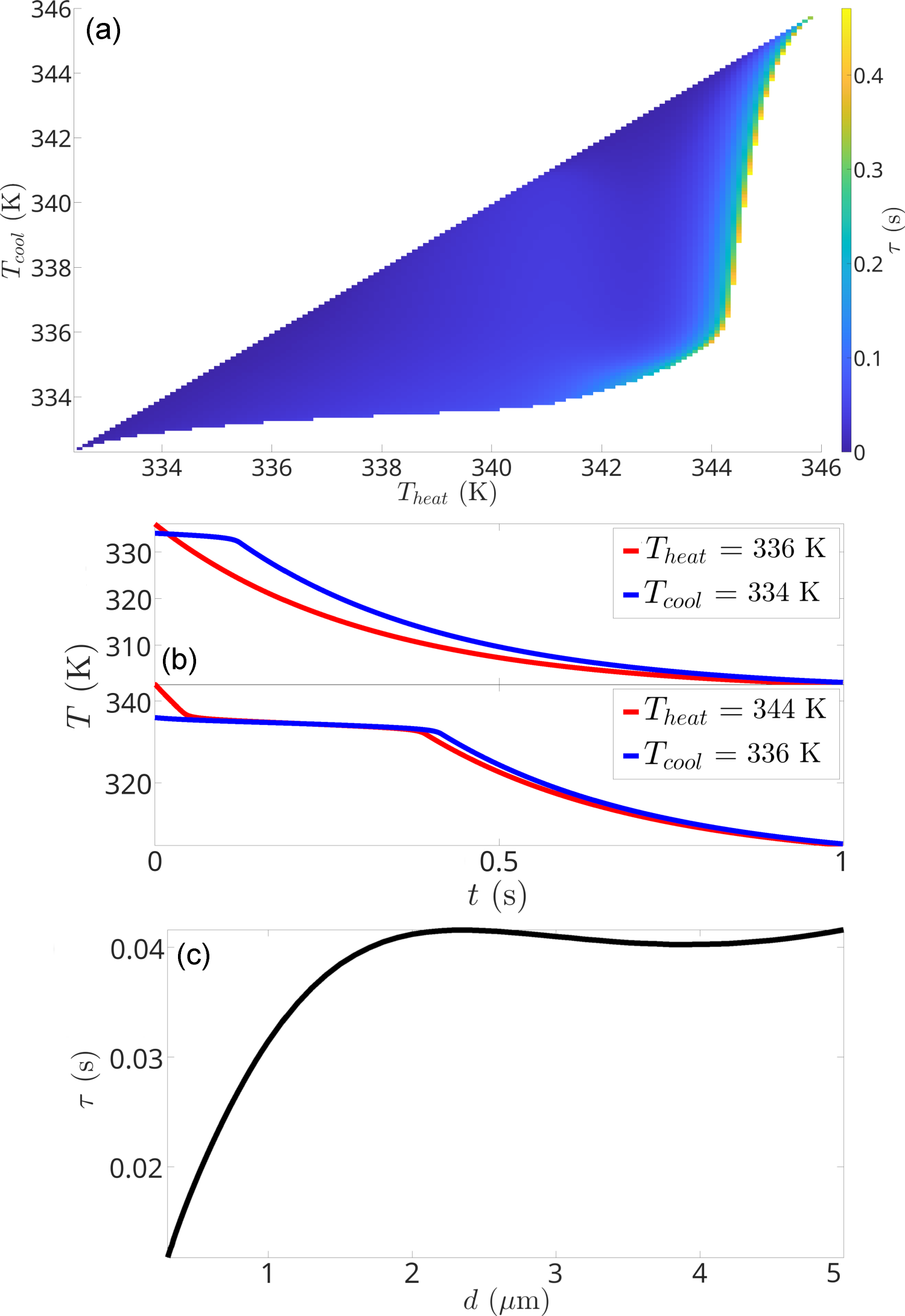}
	\caption{Mpemba effect. (a) Heatmap of the time $\tau$ at which two temperature curves intersect which were initially prepared on the heating branch $T_{\rm heat}$ and the cooling branch $T_{\rm cool}$ of the hysteresis. The VO$_2$-particle's radius is $R=100$ nm, the distance to the substrate $d=500$ nm, and the substrate's temperature $T_{\rm sub} = 300$ K. (b) Two exemplary temperature combinations for which the Mpemba effect occurs. (c) Distance dependence of $\tau$ for $T_{\rm heat} = 336$ K and $T_{\rm cool} = 334$ K.}
	\label{Fig:normME}
\end{figure}

For a VO$_2$ particle of radius $R=100$ nm placed at $d = 500$ nm above a SiC substrate at $T_{\rm sub} = 300$ K, Fig.~\ref{Fig:normME}(a) maps the conditions for the Mpemba effect. The color represents the intersection time $\tau$, where one particle is prepared on the cooling branch $T_{\rm cool}$ and another on the heating branch $T_{\rm heat}$. Although the radiative contrast $\partial \mathcal{P} / \partial f_m$ is non-negligible, which is evidenced by the exchanged power discussed in Sec.~S2 of the Supplemental Material, the characteristic features of the Mpemba effect are predominantly governed by the caloric channel. This dominance arises from the transient behavior of $C_{\rm p}$ within the phase-change regime. While the radiative power $\mathcal{P}_{{\rm sub} \leftrightarrow {\rm dip}}$ varies gradually over the transition width, the latent heat contribution introduces a massive, localized peak in $C_{\rm p}$ via the derivative $\partial f/\partial T$. This caloric peak acts as a strong thermal buffer that dynamically alters the cooling trajectories. Consequently, the distinct features of the observed Mpemba effect can be almost exclusively attributed to this latent-heat-induced modulation of the heat capacity.

Fig.~\ref{Fig:normME}(b) shows two exemplary temperature curves exhibiting early and later intersection times $\tau$. In both cases, the latent heat delays the temperature decay, generating a visible plateau. This plateau emerges in the regime of rapid change in $f_{m/d}$ after entering the hysteresis' slopes illustrated in Fig.~\ref{Fig:Sketch}. The case for $T_{\rm heat} = 344$ K and $T_{\rm cool} = 336$ K shows that for certain temperatures within the transition regime (e.g. $T_{\rm heat} = 345$ K), two curves can miss each other if the intersection does not coincide with the phase of latent heat release. If $T_{\rm heat}$ has not yet reached the steep slope of the heating branch, there is no delay due to latent heat because $\partial f / \partial T \approx 0$. This enables the initially hotter curve to ``overtake'' the $T_{\rm cool}$ curve while the latter experiences the latent heat delay, as illustrated for $T_{\rm heat} = 336$ K and $T_{\rm cool} = 334$ K in Fig.~\ref{Fig:normME}(b). Note that the here implemented Temperature First-Order Reversal Curve (T-FORC) model (see Sec.~S1 in the Supplemental Material) prevents a steep slope of $f_{m/d}$ at early times, physically accounting for the kinetic delay in domain reorganization upon thermal reversal. This explains why the region $T_{\rm heat} < T_{\rm cool}$ does not yield a Mpemba effect when preparing $T_{\rm heat}$ directly on the heating slope. Conversely, near $T_{\rm heat/cool} \approx 332$ K, the curves can miss each other if the cooling branch is not yet steep enough at $T_{\rm cool}$. If $T_{\rm heat} < T_{\rm cool}$ is prepared at those lower temperatures, latent heat will always delay the decay of $T_{\rm cool}$, generally preventing an intersection. 

Additionally, Fig.~\ref{Fig:normME}(c) depicts the distance dependence of $\tau$ for $T_{\rm heat} = 336$ K and $T_{\rm cool} = 334$ K, highlighting the role of near-field coupling. Shorter distances accelerate the temperature decay. For $d > 2$ \textmu m, the near-field coupling becomes negligible, and $\tau$ oscillates around a far-field asymptote of $\tau_{\infty} \approx 0.04$ s. These oscillations stem from the propagating waves embedded in the Local Density of States (LDOS), which dominates once the evanescent contributions can be neglected. Similar oscillatory relaxation behaviors in the far field have been reported in Ref.~\cite{Tschikin2012} for SiC and gold configurations, where they were analytically evaluated using the stationary phase method. Note that changing the distance primarily shifts the intersection time $\tau$ but leaves the boundaries of the Mpemba phase space in Fig.~\ref{Fig:normME}(a) almost unchanged.

It is important to emphasize that this radiative Mpemba effect is not a simple thermal delay. Rather, it represents a selective trajectory crossing induced by the strong, localized peak in the specific heat capacity $C_{\rm p}$ due to latent heat release during path-dependent domain reorganization. While the near-field radiative channel dictates the global relaxation time scale $\tau$, the caloric peak acts as a dynamic buffer that allows initially hotter trajectories to overtake cooler ones.

\begin{figure}
	\centering
	\includegraphics[width=0.48\textwidth]{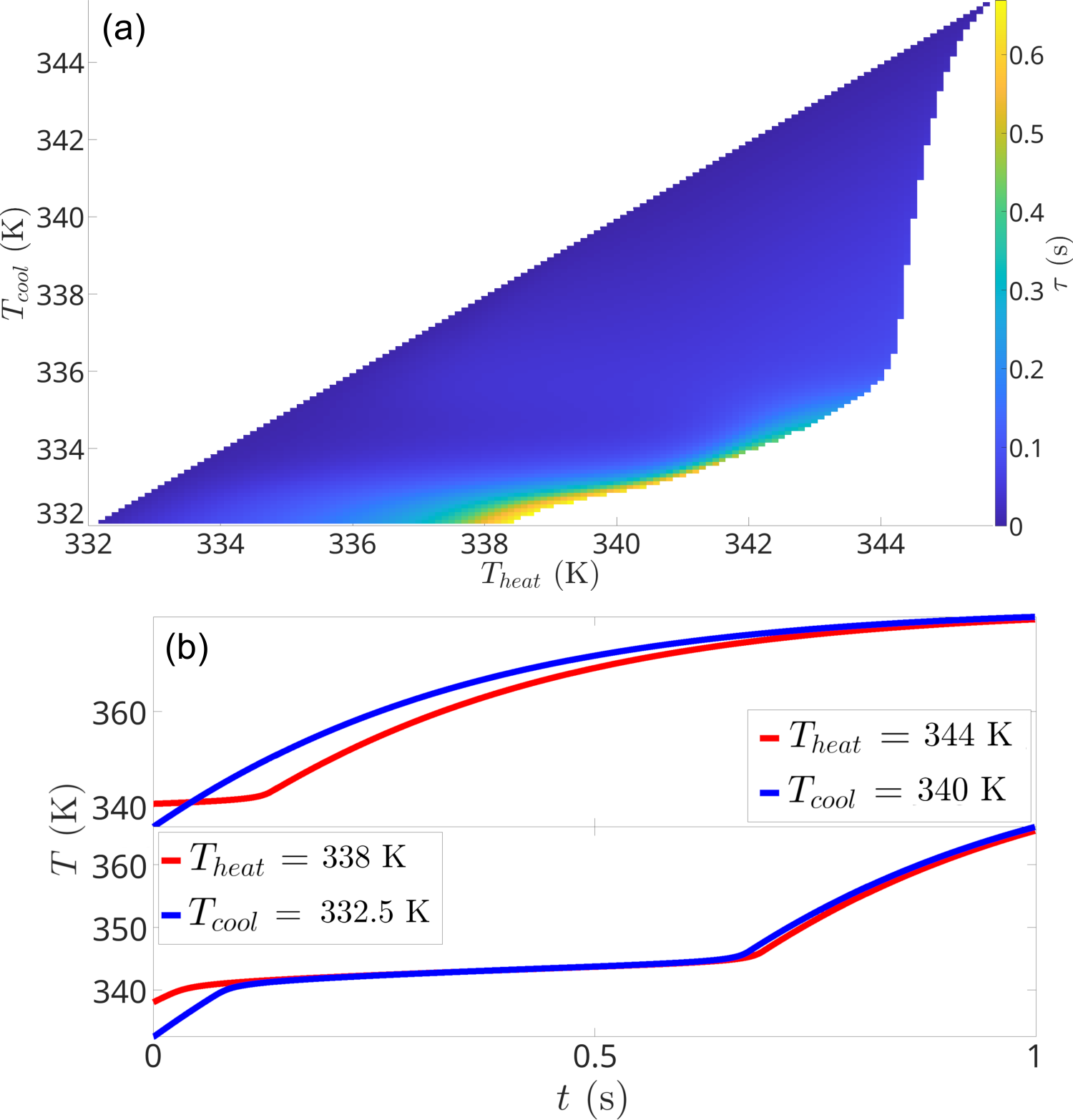}
	\caption{Inverse Mpemba effect. (a) Heatmap identical to Fig.~\ref{Fig:normME} but with the substrate temperature $T_{\rm sub} = 378$ K causing an inverse Mpemba effect. (b) Two exemplary temperature combinations for which the inverse Mpemba effect occurs.}
	\label{Fig:invME}
\end{figure}

Fig.~\ref{Fig:invME}(a) illustrates the conditions for the inverse Mpemba effect, obtained by setting the substrate temperature to $T_{\rm sub} = 378$ K to induce heating. While the overall topology of the heatmap resembles that of the ordinary Mpemba effect, the maximum intersection times $\tau$ shift toward lower initial temperatures. This shift reflects the fundamental inversion of the hysteresis pathways. During heating, the structural evolution along the primary heating branch remains unaltered, acting as a fixed reference where latent heat absorption occurs at higher temperatures. In contrast, a particle prepared on the cooling branch at $T_{\rm cool}$ is now forced onto an ascending T-FORC pathway upon reversal. Because this T-FORC trajectory starts flat, the cooler particle is temporarily shielded from latent heat suppression due to $\partial f / \partial T \approx 0$. Consequently, an intersection either occurs when $T_{\rm heat}$ is delayed by latent heat at an early stage, or while both curves are simultaneously delayed by latent heat, which yields the maximum $\tau$ values. Both regimes are exemplified in Fig.~\ref{Fig:invME}(b).

To discuss the possibility of a passive Mpemba effect when exchanging the materials of the particle and the substrate, the semi-infinite substrate is replaced by a thin film of thickness $t_{\rm sub} = 100$ nm. This prevents the necessity of accounting for the anisotropy of bulk VO$_2$, which otherwise manifests for large thicknesses \cite{Qazilbash2009, vanZwol2011}. Therefore, the Fresnel amplitude reflection coefficients $r_{s/p}$ of a semi-infinite half space described in Sec.~S1 of the Supplemental Material are replaced by the Airy-tipe expressions \cite{Chew1995}
\begin{align}
r_{s/p, {\rm slab}} & = \frac{r_{s/p} (1 - e^{2 \text{i} k_{z, {\rm sub}} t_{\rm sub}})}{1 - r_{s/p}^2 e^{2 \text{i} k_{z, {\rm sub}} t_{\rm sub}}} .
\end{align}

\begin{figure}
	\centering
	\includegraphics[width=0.48\textwidth]{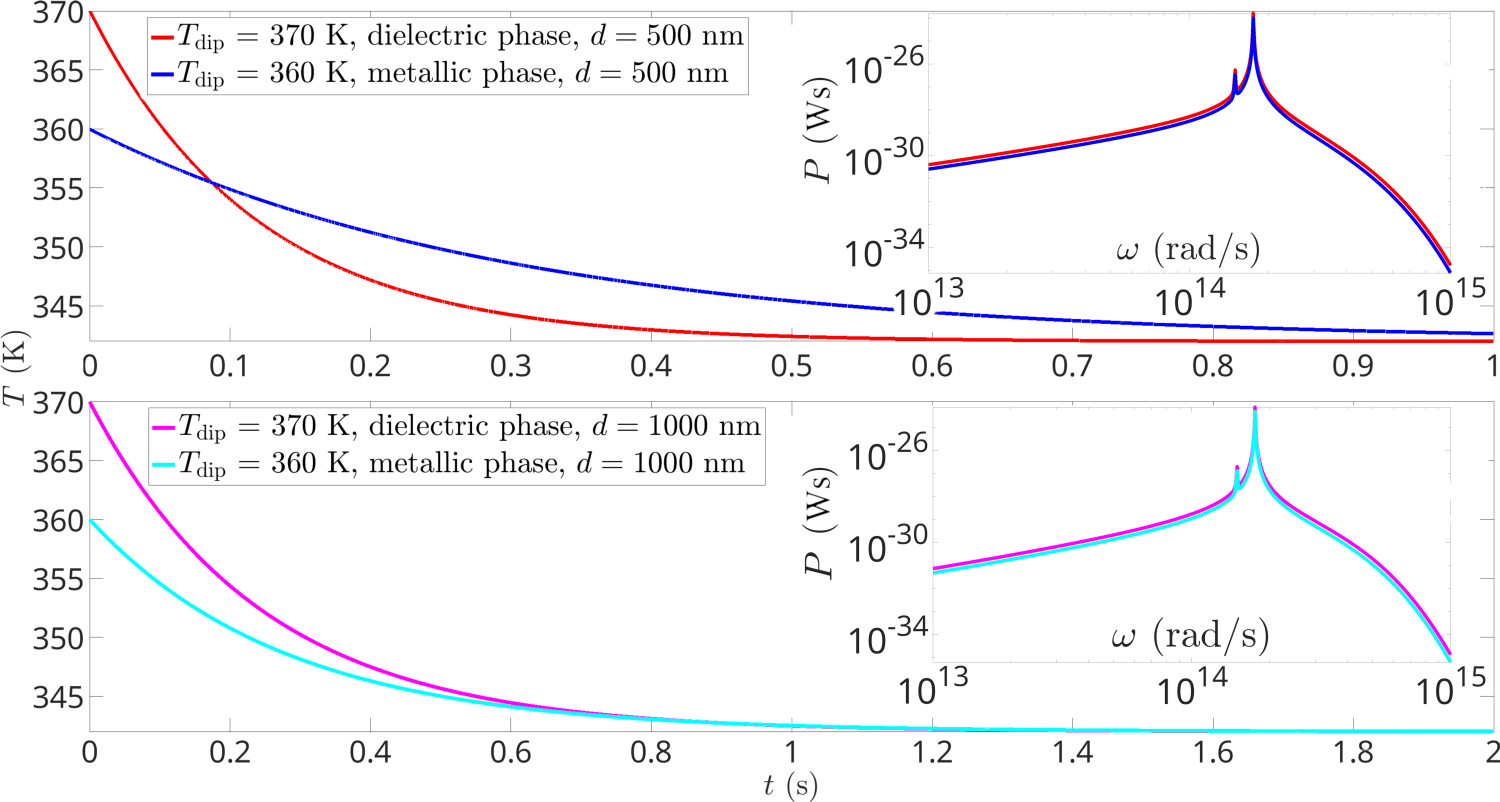}
	\caption{Passive Mpemba effect. Temperature evolution of a $R=100$ nm SiC particle at $d=500$ nm (red and blue) and $d=1$ \textmu m (magenta and cyan) above a VO$_2$ substrate of thickness $t=100$ nm. The initial temperature of the particle above the substrate prepared in the dielectric phase is $T_{\rm dip} = 370$ K (red and magenta), whereas $T_{\rm dip} = 360$ K (blue and cyan) corresponds to the case where the substrate is prepared in the metallic phase. Insets: Corresponding spectral power for each initial temperature.}
	\label{Fig:passME}
\end{figure}

Now, Fig.~\ref{Fig:passME} shows the time evolution for two initial temperatures $T_{\rm dip} = 360$ K and $T_{\rm dip} = 370$ K while keeping the VO$_2$ film at $T_{\rm sub} = 342$ K, once prepared in the dielectric phase and once in the metallic phase. The distance is varied between $d=500$ nm and $d=1$ \textmu m. For $d=500$ nm, the initially hotter SiC particle cools down faster than the initially colder one, intersecting it at $\tau \approx 0.08$ s. The reason for this is shown in the insets, which depict the exchanged power for both phases. In the metallic phase of the VO$_2$ film, the SiC particle cannot couple as effectively to the substrate as in the dielectric phase due to higher reflection. Therefore, the heat exchange takes much longer in the metallic phase than in the dielectric one, giving rise to the passive Mpemba effect. When, however, increasing the distance to $d=1$ \textmu m, this coupling becomes weaker in the dielectric phase so that the passive Mpemba effect effectively vanishes and cannot be distinguished from numerical noise anymore. This directly proves that the passive Mpemba effect only occurs in the near field.

\section{Conclusion} \label{Ch:Conc}

In conclusion, this work has translated the anomalous relaxation of the Mpemba effect to thermal radiative transport. By exploiting the coupling between near-field radiation and the hysteresis pathways of phase-change materials, it is demonstrated that an initially hotter system can cool down faster than an initially colder one. Beyond establishing this baseline phenomenon, this study reveals that the radiative Mpemba effect is not a simple thermal delay, but rather a dynamic interplay where latent heat acts as a caloric buffer, controlling trajectory crossings while near-field photon exchange dictates the overall relaxation timescale. A generalized analytical condition for this radiative Mpemba effect has been derived, showing that its emergence is dictated by the interplay between radiative power and specific heat capacity derivatives. Through numerical simulations of a VO$_2$ nanoparticle interacting with a SiC substrate, the Mpemba phase space has been mapped, verifying the three distinct manifestations of the ordinary, inverse, and passive Mpemba effects. Crucially, the passive variant highlights that the anomaly can be triggered solely via external memory storage in the substrate reflection, proving that the relaxing system itself does not require intrinsic hysteresis. Furthermore, the breakdown of this passive variant at larger separation distances confirms its strictly near-field nature, demonstrating geometric tunability as the system transitions from the near- to the far-field regime. 

The chosen setup matches readily accessible experimental configurations, facilitating laboratory verification via pump-probe microscopy \cite{Wagner2014, Doenges2016}. The magnitude of the effect could potentially be enhanced by using symmetric VO$_2$-VO$_2$ configurations instead, maximizing the phase-change radiative contrast due to ideal resonant frequency-matching of the SPhPs in the dielectric phase \cite{vanZwol2011}. Additionally, the underlying hysteresis loop can be precisely tailored to specific operating temperatures, for instance, by doping with tungsten \cite{Jin1995, Tan2012}. Consequently, these findings open up novel avenues for exploiting material history and structural memory in radiative heat transfer, establishing a versatile cornerstone for advanced dynamic thermal management on the nanoscale.

%%%%%%%%%%%%%%%%%%%%%%%%%%%%%%%%%%%%%%%%%%%%%%%%%%%%%%%%%%%%%%%%%%%%%%%%%%%%%%%%
%
% Acknowledgement
%
%%%%%%%%%%%%%%%%%%%%%%%%%%%%%%%%%%%%%%%%%%%%%%%%%%%%%%%%%%%%%%%%%%%%%%%%%%%%%%%%
\section*{Acknowledgements}

The author acknowledges financial support by Deutsche Forschungsgemeinschaft under project number 570757245 and fruitful discussions with Philippe Ben-Abdallah on the Mpemba condition. Simulations were conducted on the the HPC cluster ROSA funded by the Deutsche Forschungsgemeinschaft under INST 184/225-1 FUGG.

\bibliography{References.bib}

\end{document}

\title{Supplemental Material: Hysteresis-Driven Radiative Mpemba Effect in Phase-Change Nanostructures}

\author{F. Herz}
\affiliation{Institut f\"{u}r Physik, Carl von Ossietzky Universit\"{a}t, 26111, Oldenburg, Germany}
\email{florian.herz@uol.de}

\date{\today}

\maketitle

\setcounter{equation}{0}
\setcounter{figure}{0}
\setcounter{table}{0}
\setcounter{page}{1}

\makeatletter
\renewcommand{\theequation}{S\arabic{equation}}
\renewcommand{\thefigure}{S\arabic{figure}}
\renewcommand{\thetable}{S\arabic{table}}
\renewcommand{\thepage}{S\arabic{page}}

%%%%%%%%%%%%%%%%%%%%%%%%%%%%%%%%%%%%%%%%%%%%%%%%%%%%%%%%%%%%%%%%%%%%%%%%%%%%%%%%
%
% Introduction
%
%%%%%%%%%%%%%%%%%%%%%%%%%%%%%%%%%%%%%%%%%%%%%%%%%%%%%%%%%%%%%%%%%%%%%%%%%%%%%%%%

\section{S1. Thermal radiative transport}

The cooling rate of the VO$_2$ particle is given by the heat balance equation \cite{Tschikin2012}
\begin{equation}
\frac{\text{d} T_{\rm dip}}{\text{d} t} = - \frac{\mathcal{P}_{{\rm sub} \leftrightarrow {\rm dip}} (T_{\rm dip})}{\rho C_{\rm p} (T_{\rm dip}) V}
\end{equation}
which is valid if the characteristic time scale for the temperature homogenization within the particle is much faster than the thermal relaxation time due to thermal radiation. Since this characteristic time scale is in the range of $10^{-8}$ s while the thermal relaxation due to thermal radiation takes place in the range of $10^{-3}$ s, this assumption is fulfilled. Here, $T_\text{dip}$ describes the VO$_2$ particle's temperature, $t$ the time, $\rho$ its mass density, $C_\text{p}$ its specific heat capacity, $V$ its volume, and $\mathcal{P}_{{\rm sub} \leftrightarrow {\rm dip}} (T_{\rm dip})$ the exchanged power between the particle and the underlying SiC substrate. 

The exchanged power can be expressed as \cite{Tai1971, Sipe1987, Mulet2001, Rousseau2009, Huth2010, Kallel2017, Herz2018, Herz2021, Herz2022_2}
\begin{equation}
\mathcal{P}_{{\rm sub} \leftrightarrow {\rm dip}} (T_{\rm dip}) = \sum_{i=\text{E,H}} \int_0^\infty \text{d} \omega \, 2 \omega \text{Im} (\alpha_i) D_i(\omega, d) \Delta \Theta(\omega, T_{\rm dip}, T_{\rm sub})
\label{Eq:Pow}
\end{equation}
with the electric and magnetic polarizabilities $\alpha_\text{E/H}$, their Local Density of States (LDOS) $D_\text{E/H}$ and the Bose-Einstein energy difference distributions
\begin{equation}
\Delta \Theta(\omega, T_\text{dip}, T_\text{sub}) = \hbar \omega \left[ \frac{1}{e^{\frac{\hbar \omega}{k_B T_\text{dip}}} - 1}  - \frac{1}{e^{\frac{\hbar \omega}{k_B T_\text{sub}}} - 1} \right].
\end{equation}
The LDOS between a spherical nanoparticle and a planar, semi-infinite substrate reads as
\begin{align}
D_\text{E/H}(\omega, d) & = \frac{\omega^2}{2 \pi c^3} \int \frac{\text{d}^2 k_\perp}{(2 \pi)^2} \frac{e^{-2 \text{Im}(k_z) d}}{|k_z|^2 k_0} \biggl[ \text{Re}(k_z) \left( 2 + \text{Re} \left( r_{s/p} e^{2 \text{i} k_z d} \right) + \frac{2 k_\perp^2 - k_0^2}{k_0^2} \text{Re} \left( r_{p/s} e^{2 \text{i} k_z d} \right) \right) \notag \\
& \quad + \text{Im} \left( k_z \right) \left( \text{Im} (r_{s/p}) + \frac{2 k_\perp^2 - k_0^2}{k_0^2} \text{Im}(r_{p/s}) \right) \biggr]
\end{align}
while using the Fresnel amplitude reflection coefficients
\begin{align}
r_s & = \frac{k_z - k_{z, \text{sub}}}{k_z + k_{z, \text{sub}}} , 
\label{eq:rs} \\
r_p & = \frac{\varepsilon_\text{sub} k_z - k_{z, \text{sub}}}{\varepsilon_\text{sub} k_z + k_{z, \text{sub}}}
\label{eq:rp}
\end{align}
and the abbreviations
\begin{align}
k_z & = \sqrt{k_0^2 - k_\perp^2} , \\
k_{z, \text{sub}} & = \sqrt{\varepsilon_\text{sub} k_0^2 - k_\perp^2} .
\end{align}
Note that the real parts of both the electric and magnetic parts of the LDOS contain contributions of propagating waves and evanescent waves which are characterized by $k_\perp < k_0$ and $k_\perp > k_0$, dominating the far field or near field, respectively. The magnetic and electric parts of the LDOS can be changed into each other by exchanging the Fresnel amplitude reflection coefficients $r_s \leftrightarrow r_p$. The electric and magnetic polarizabilities are given by the Mie coefficients in the long-wavelength limit $x = k_0 R \ll 1$
\begin{align}
\alpha_\text{E} & = 2 \pi R^3 \frac{(2 \varepsilon_\text{dip} + 1) \left[ \sin(y) - y \cos(y) \right] - y^2 \sin(y)}{(\varepsilon_\text{dip} - 1) \left[ \sin(y) - y \cos(y) \right] + y^2 \sin(y)} , \\
\alpha_\text{H} & = \frac{\pi R^3}{3} \left(\frac{x^2 - 6}{y^2} \left[ y^2 + 3 y \cot(y) - 3 \right] - \frac{2 x^2}{5} \right) ,
\end{align}
which is fulfilled for the radius of $R=100$ nm considered here, while $y = \sqrt{\varepsilon_\text{dip}} x$ \cite{Bohren1983}. For the permittivities of the two phases, a Drude-Lorentz model for the dielectric phase 
\begin{equation}
\varepsilon_d(\omega) = \varepsilon_{\infty} + \sum_{i=1}^n \frac{S_i}{1 - \frac{\omega^2}{\omega_i^2} - \text{i} \Gamma \frac{\omega}{\omega_i}}
\end{equation}
is used and a Drude model for the metallic phase
\begin{equation}
\varepsilon_m(\omega) = - \frac{\omega_p^2 \varepsilon_{\infty}}{\omega^2 + \text{i} \omega \omega_c}.
\end{equation}
For the SiC substrate, the same Drude-Lorentz model is used while taking $n=1$. The optical parameters for VO$_2$ and SiC are taken from Ref.~\cite{Barker1966} and Ref.~\cite{Bohren1983}.

In the transition regime, VO$_2$ switches from a dielectric phase into a metallic phase with rising temperatures. Since this behavior is different for the heating and cooling processes, it is described by a hysteresis with respect to the metallic volume fraction $f_m$ and the dielectric volume fraction $f_d$. These volume fractions also define the permittivity of VO$_2$ by means of Bruggeman's symmetric effective medium model \cite{Goodenough1971, Choi1996, Jepsen2006}
\begin{equation}
0 = f_m \frac{\varepsilon_m - \varepsilon_\text{dip}}{\varepsilon_m + 2 \varepsilon_\text{dip}} + f_d \frac{\varepsilon_d - \varepsilon_\text{dip}}{\varepsilon_d + 2 \varepsilon_\text{dip}}
\end{equation}
assuming spherically shaped nanoparticles. When systems undergoing hysteresis are prepared within the transition regime, for instance, on the heating branch, and forced to cool down, they do not simply retrace the heating branch downwards. Instead, the classical approach to describe these curves between the outer branches of the hysteresis loop relies on Preisach density distributions \cite{Mayergoyz2003}, typically evaluated via Temperature First-Order Reversal Curves (T-FORCs) to match experimental observations \cite{Tanasa2005}. In this work, a simplified, analytical version of this approach is employed to avoid computationally heavy integrations associated with full Preisach density distributions, utilizing the following model instead
\begin{align}
f_{\text{forc}, \uparrow/\downarrow}(T) & = f_{m/d}(T) + \left( f_{\text{forc}, \uparrow/\downarrow}(T_0) - f_{m/d}(T_0) \right) \frac{f_{m/d}(T)}{f_{m/d}(T_0)} ,
\end{align}
where the upper index pair ($\uparrow$ and $m$) corresponds to the metallic volume fraction during the reversal onto a heating pathway, while the lower index pair ($\downarrow$ and $d$) describes the dielectric volume fraction during the reversal onto a cooling pathway. Note that this model is only used if the system is prepared within the transition regime. If it is initially prepared in a temperature range above or below this regime, it will follow the outer hysteresis branches which are defined by $f_m$ and $f_d$ and modeled by 
\begin{align}
f_m (T) & = - \frac{T}{U_m} W_0 \left( -\frac{U_m}{2 T} e^{-\frac{U_m}{T}} \text{erfc} \left( \frac{T_{c,m} - T}{\sqrt{2} \Delta T_m}\right) \right),  \\
f_d (T) & = \frac{T}{U_d} W_0 \left( \frac{U_d}{2 T} e^{-\frac{U_d}{T}} \text{erfc} \left( \frac{T - T_{c,d}}{\sqrt{2} \Delta T_d}\right) \right)
\end{align}
with $U$ being the driving force responsible for the hysteresis of VO$_2$, $\Delta T$ being the specific temperature width describing the range of the slope, $T_c$ denoting the critical temperature, $W_0$ the Lambert W function, and erfc the complementary error function. Here, I used $U_m = 233.2$ K, $T_{c,m} = 342.5$ K, and $\Delta T_m = 1$ K as well as $U_d = 10.1$ K, $T_{c,m} = 334.5$ K, and $\Delta T_d = 0.8$ K \cite{Ordonez-Miranda2018}.

The specific heat capacity of VO$_2$ can be modelled by two components, namely the Debye heat capacity $C_D$ and the contribution of the phase transition while neglecting the thermal expansion which was shown experimentally to be much smaller than the other two \cite{Chandrashekhar1973}, so that \cite{Kawakubo1964, Zhong2011, Ordonez-Miranda2018}
\begin{equation}
C_{p, \uparrow/\downarrow}(T) = C_D(T) + L \frac{\text{d} f_{{\rm forc} \uparrow/\downarrow}}{\text{d} T}
\end{equation}
holds. Here, $L = 4268 \, \frac{\text{J}}{\text{mol}}$ denotes the latent heat \cite{Berglund1969}. The Debye model 
\begin{equation}
C_D(T) = 9 n_A \frac{N_A k_B}{M} \frac{T}{\Theta_D} \int_0^{\Theta_D/T} \text{d} x \frac{x^4 e^{x}}{\left( e^{x} - 1 \right)^2}
\end{equation}
uses Avogadro's constant $N_A = 6.022 \times 10^{23}$ mol$^{-1}$, the molar mass $M$, the number of atoms per unit cell $n_a$, and the Debye temperature $\Theta_D$. For VO$_2$, $M = 0.083 \, \frac{\text{kg}}{\text{mol}}$, $n_A = 3$, and $\Theta_D = 750$ K are used \cite{Ordonez-Miranda2018}. When switching to the passive Mpemba effect, the material parameters for the specific heat capacity of SiC are $M = 0.0401 \, \frac{\text{kg}}{\text{mol}}$, $n_A = 2$, and $\Theta_D = 1100$ K \cite{Tschikin2012}.

\section{S2. Spectral power}

\begin{figure}
	\centering
	\includegraphics[width=0.9\textwidth]{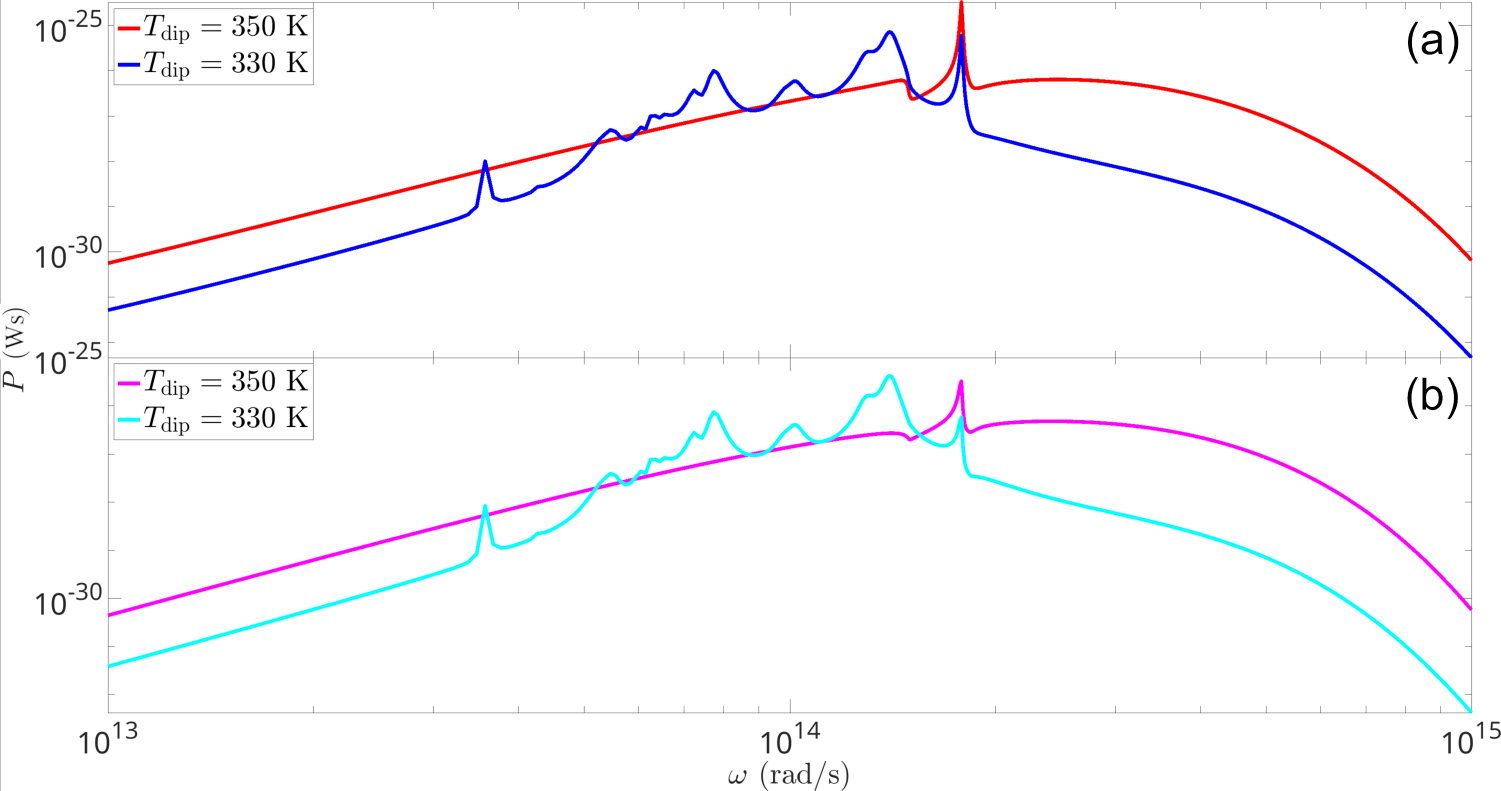}
	\caption{Spectral power representing the integrand of Eq.~\eqref{Eq:Pow} in the metallic phase ($T_{\rm dip} = 350$ K) and in the dielectric phase ($T_{\rm dip} = 330$ K) for the center-to-surface distances $d=500$ nm (a) and $d=1$ \textmu m (b).}
	\label{Fig:Pow}
\end{figure}

According to Eq.~\eqref{Eq:Pow}, the spectral power exchanged between the VO$_2$ particle and the SiC substrate is expressed as
\begin{equation}
\mathcal{P} = \sum_{i=\text{E,H}} 2 \omega \text{Im} (\alpha_i) D_i(\omega, d) \Delta \Theta(\omega, T_{\rm dip}, T_{\rm sub}) .
\label{Eq:spec_Pow}
\end{equation}
To demonstrate the influence of the two different phases, Fig.~\ref{Fig:Pow} shows the results of Eq.~\eqref{Eq:spec_Pow} for two different temperatures, namely $T_{\rm dip} = 330$ K corresponding to the dielectric phase and $T_{\rm dip} = 350$ K corresponding to the metallic phase, evaluated at two different distances $d=500$ nm and $d=1$ \textmu m. In all cases, the spectral peak at $\omega = 1.787 \times 10^{14}$ rad/s corresponding to the substrate's surface phonon polaritons (SPhPs) is clearly visible. In the dielectric phase, a strong peak at $\omega = 1.398 \times 10^{14}$ rad/s also dominates the spectrum, which highlights the localized surface phonon polaritons (LSPhPs) of the spherical VO$_2$ particle. In the dielectric phase, both peaks are visible, and only for larger distances of $d>500$ nm, the LSPhP peak remains as the only characteristic peak due to the reduced coupling to the SiC surface. In the metallic phase, the VO$_2$ particle possesses no intrinsic localized resonances, so that the SiC-SPhP peak simply superimposes onto the broad spectrum but also diminishes for larger distances. Although the peaks in the dielectric phase can be larger than in the metallic phase, the frequency integration shows that, due to the broader spectrum of the metallic phase, the total exchanged power in the metallic phase is approximately twice that of the dielectric phase.

\bibliography{References.bib}